Electron g-factor determined for quantum dot circuit fabricated from (110)-oriented GaAs quantum well

T. Nakagawa,[1] S. Lamoureux,[2] T. Fujita,[1,3] J. Ritzmann,[4] A. Ludwig,[4] A. D. Wieck,[4] A. Oiwa,[1,3,5,a)] M. Korkusinski,[2] A. Sachrajda,[2] D. G. Austing,[2] and L. Gaudreau[2,b)]

[1]SANKEN, Osaka University, 8-1 Mihogaoka, Ibaraki-shi, Osaka 567-0047, Japan

[2]Emerging Technology Division, National Research Council, Ottawa, K1A0R6, Canada

[3]Center for Quantum Information and Quantum Biology (QIQB), Osaka University, Osaka 565-0871, Japan

[4]Lehrstuhl für Angewandte Festkörperphysik, Ruhr-Universität Bochum, Universitätsstraße 150, Gebäude NB, D-44780 Bochum, Germany

[5]Center for Spintronics Research Network (CSRN), Graduate School of Engineering Science, Osaka University, Osaka 560-8531, Japan

a)Author to whom correspondence should be addressed: oiwa@sanken.osaka-u.ac.jp

b)Author to whom correspondence should be addressed: Louis.Gaudreau@nrc-cnrc.gc.ca

(Dated: October 5, 2021)

ABSTRACT:

The choice of substrate orientation for semiconductor quantum dot circuits offers opportunities for tailoring spintronic properties such as g-factors for specific functionality. In this letter, we demonstrate the operation of a few-electron double quantum dot circuit fabricated from a (110)-oriented GaAs quantum well. We estimate the in-plane electron g-factor from the profile of the enhanced inter-dot tunneling (leakage) current near zero magnetic field. Spin-blockade due to Pauli exclusion can block inter-dot tunneling. However, this blockade becomes inactive due to hyperfine interaction mediated spin flip-flop processes between electron spin states and the nuclear spin of the host material. The g-factor of absolute value ~0.1 found for a magnetic field parallel to the direction [$\bar{1}10$], is approximately a factor of four lower than that for comparable circuits fabricated from material grown on widely-employed standard (001) GaAs substrates, and is in line with reported values determined by purely optical means for quantum well structures grown on (110) GaAs substrates.

Quantum spintronic properties offer a valuable and diverse resource for numerous quantum nano-technologies: see for example the reviews in Refs. [1, 2]. An important choice at the start of a process to fabricate a semiconductor spintronic device concerns the substrate orientation. While the standard and readily available (001) substrate is widely employed, other substrate orientations may be appealing. In the context of potential spintronic applications, quantum wells (QWs) grown on (110) GaAs substrates as compared to those grown on (001) GaAs substrates have attracted interest for properties such as suppressed spin relaxation and spin relaxation anisotropy [3-13], unidirectional (out-of-plane) spin-orbit interaction and spin-helix state [14-20], electron g-factor anisotropy [21], anisotropic spin transport [22], and absence of weak anti-localization for a perfectly symmetric QW [23].

GaAs QWs grown on GaAs substrates other than (001) also offer large and highly anisotropic g-factors for holes confined in two dimensions [24]. For the case of a (001)-oriented QW, the heavy-hole g-factor is close to zero for a magnetic (B-) field applied in the plane of the QW [25-27]. In contrast, for a low symmetry growth direction such as [110], the in-plane heavy-hole g-factor is significant and anisotropic. A GaAs QW grown on a (110) GaAs substrate is potentially attractive for coherent photon-to-spin conversion [28-31]. This is because the lowest energy states in the valence band with dominant heavy-hole character may be significantly split for the Voigt measurement configuration to enable the necessary V-shaped arrangement of three levels to transcribe the photon state to the electron spin state [28-31].

In this letter, we demonstrate the operation of a few-electron double quantum dot (DQD) with nearby charge detector that is fabricated from a (110)-oriented GaAs QW structure. We estimate the in-plane electron g-factor from the Gaussian profile of the elevated leakage current near zero B-field arising from the partial lifting of Pauli spin blockade (PSB) [32-34] by electron-spin nuclear-spin mixing facilitated by the hyperfine interaction [34]. The g-factor of absolute value ~0.1 found for an in-plane B-field directed along [$\bar{1}$10] is approximately a factor of four lower than that typically encountered for comparable DQD devices fabricated from hetero-structures grown on more standard (001) GaAs substrates.

The details of the QW structure grown by molecular beam epitaxy are as follows. On the (110) GaAs substrate, first a thick 1000 Å buffer layer is grown followed by a smoothing and clean-up superlattice with 20 periods of 24 Å GaAs/ 24 Å AlAs. The subsequent epilayers consist of a 1000 Å $Al_{0.33}Ga_{0.67}As$ lower barrier, a 120 Å GaAs QW, a 250 Å $Al_{0.33}Ga_{0.67}As$ undoped upper barrier, a 350 Å $Al_{0.33}Ga_{0.67}As$ Si-doped upper barrier, and finally a 50 Å GaAs cap layer. The growth temperature is 490°C throughout the whole growth process except for one additional 120 s high temperature annealing break at a temperature of 635°C after the first 1000 Å buffer layer. All temperatures are given pyrometer readings. During the whole structure and annealing breaks, a relative high arsenic beam equivalent pressure of $1.3 \times 10^{-5}$ torr is maintained with the As-cracker zone kept at 700°C, i.e., mainly keeping the $As_4$-molecules uncracked. For our preliminary investigation of a (110)-oriented GaAs QW structure for gated QD devices, the choice of a 12-nm wide QW, inspired by the work of Hübner *et al*. [21], offers a finite electron in-plane g-factor although of a smaller absolute value compared to that for a (001)-orientated QW of the same width. The finite g-factor allows spin manipulation, although we note that a reduction in the well width to ~7.5 nm can drive the in-plane g-factor to essentially zero [21, 38]: a feature attractive for coherent photon-to-spin conversion [28-31].

To characterize the two-dimensional electron gas confined in the (110)-oriented GaAs QW quantum Hall measurements are performed with a standard Hall bar. Figure 1 shows the longitudinal resistivity ($\rho_{xx}$) and transverse resistivity ($\rho_{xy}$) measured in the dark at 1.5 K. The carrier density and mobility respectively are determined to be $9.3 \times 10^{10}$ cm$^{-2}$ and $8.5 \times 10^{4}$ cm$^{2}$/Vs.

Figure 2(a) shows a scanning electron micrograph of the DQD device highlighting just the active metallic gates on the surface of the (110)-oriented GaAs/AlGaAs QW structure. A global top gate is also employed (not shown in image). Voltages $V_L$ and $V_R$ respectively on the left (L) and right (R) plunger gates regulate the number of electrons M and N on the left and right QDs. The current, $I_{DQD}$, flowing through the DQD, in response to a source-drain bias voltage, $V_{SD}$, and the current, $I_{CS}$, flowing through the nearby QPC charge sensor are measured. An in-plane B-field is applied parallel to the [$\bar{1}$10] direction (the DQD is aligned along the $[00\bar{1}]$ direction). All measurements on the DQD device are performed in a dilution refrigerator and the electron temperature is ~135 mK.

For the DQD investigated, we are able to enter the few-electron regime, but are impacted by potential fluctuations leading to the formation of incidental QDs that couple to the DQD of interest. We believe this non-ideal potential landscape reflects the material grown on the non-standard (110) substrate. In particular, the left QD region splits into a larger QD and a smaller QD: see cartoon inset in Fig. 2(b). Because of this, the full stability diagram shown in Fig. 2(b) is quite complex, and we could not study in isolation the region of the stability diagram with the minimal configuration of one permanently trapped electron in either the left QD region or the right QD region necessary to perform PSB measurements in the limit of the fewest electrons [32, 34]. Instead we focus on distinct transport triangles (that do not overlap with transport triangles from other pairs of triple points at high bias) showing clear signatures of PSB with respect to bias polarity for which the effective DQD system consists of the larger QD on the left side, containing a few electrons (the exact number is not known since gate leakage starts to occur before the dot could be emptied), and the right QD containing a known number of electrons. The smaller QD on the left side contains one electron and is a "spectator" playing no role in the following discussions. Figure 2(c) shows the charge detection signal, $dI_{CS}/dV_R$, and Fig. 2(d) shows the simultaneously measured transport signal, $I_{DQD}$, as a function of $V_L$ and $V_R$ in the vicinity of the two overlapping triangles from the same pair of triple points of interest at 0 T. Here $V_{SD}$=-0.5 mV: for negative bias polarity the electrons flow from the left side to right side of the DQD. As we will discuss soon, this is the direction for which spin-blockade is not seen near zero field. From the full stability diagram, N for the right QD is determined to be two, and M for the larger QD on the left side, hereafter called the "left QD" for simplicity, is inferred to be an odd number: see Fig. 2(c) the charge states in various regions relevant to our discussion are given in terms of M and N.

Figure 3 shows the same transport triangles as those in Fig. 2., although now for both bias polarities. In forward bias ($V_{SD}$=+0.5 mV), the current feature at the base region of the triangles (indicated by the asterisk) is clear at 0 T [panel (a)], but is diminished when a B-field is applied [see panel (b) for the case of B=1 T, and panel (c) for the case of B=2 T]. Panels (d) and (e) respectively show the corresponding transport triangles in reverse bias ($V_{SD}$=-0.5 mV) at 0 T and 4 T. The current feature along the base of the triangle is seen not only for 0 T but also for all finite fields. We attribute the general suppression of the current feature at the base of the triangles at finite field in forward bias but not in reverse bias to the PSB effect [33, 34, 36]. The lifting of spin blockade in forward bias at (and in the vicinity of) zero field due to mixing with nuclear spin is expected and we examine this further below [36]. Accounting for inert core singlet states formed from electron pairs occupying single-particle levels in each QD [33, 34], we deduce that in forward bias, neglecting spin, the effective electron [hole] cycle (1,0)→(1,1)→(2,0)→(1,0) [(2,1)→(1,1)→(2,0)→(2,1)] is energetically allowed: the numbers in the parentheses indicate the valence electrons on the left and right QDs. Including spin, however, the step (1,1)→(2,0) is blockaded when sooner-or-later the (1,1) triplet state becomes occupied [32-34, 36]. The current is suppressed if mechanisms capable of flipping spin are inoperative, or higher lying single-particle states in the down-stream QD become available. In contrast, in reverse bias, the effective electron [hole] cycle (1,0)→(2,0)→(1,1)→(1,0) [(2,1)→(2,0)→(1,1)→(2,1)] operates freely.

We now take a closer look at the behavior of the current flowing through the DQD in forward bias in the vicinity of zero field. Figure 4(a) shows the detuning dependence of $I_{DQD}$ along a section through the transport triangles (perpendicular to the base of the triangles) at $V_{SD}$=+0.5 mV for a B-field stepped through zero field. The elevated leakage current near 0 T signifying the lifting of spin blockade is a hallmark of mixing with nuclear spin [34, 36, 37]. Through the hyperfine interaction an electron spin may "flip-flop" with a nuclear spin located in one of the QDs. By this process the blockaded (1,1) triplet can be converted to a (1,1) singlet and current can flow [35]. Figure 4(b) shows a typical $I_{DQD}$ versus B cross-section through the plot in panel (a), as indicated by the dashed line, in the inelastic tunneling regime at a detuning corresponding to $V_R$=-0.7463 V. The characteristic peak shape was first discussed in Ref. [36]: see also Ref. [39]. We note that the inverse dependence whereby current is suppressed at zero field but recovers at finite field is expected when another mechanism capable of flipping electron spin, namely spin-orbit interaction, is operative: see for example Ref. [40]. In GaAs where spin-orbit interaction for electrons is weak, compared to for example InAs, we expect the influence of nuclear spin to be more significant near zero field: see also Refs. [41, 42] for further detailed discussions. We note that in our experiments here, where the DQD is aligned along the $[00\bar{1}]$ direction and the B-field is applied parallel to the $[\bar{1}10]$ direction, for electrons tunneling in the direction along the DQD axis, the linear Dresselhaus spin-orbit field is zero, and since the Rashba spin-orbit field is parallel to the applied B-field, the flipping of the electron by the spin-orbit interaction is expected to be ineffective (see Refs. [19, 20]).

The fluctuating nuclear magnetic field seen by an electron spin in a QD via the hyperfine interaction is $\sim A/g^*\mu_B\sqrt{N}$, where A, $g^*$, $\mu_B$, and N respectively are the hyperfine constant (~130 μeV for GaAs that incorporates nuclear spin I=3/2 relevant for the Ga and As isotopes [43, 44]), the effective electron g-factor, the Bohr magneton, and the typical number of nuclei over which electron wave function is spread [45]. Following the model for the B-field dependence of the inelastic leakage current described in Ref. [36], further discussed in detail in the work of Jouravlev and Nazarov [46], we fit the peak shape in Fig 4(b) to a Gaussian and determine a standard deviation σ=19±1 mT. As shown in Fig. 4(c), the value of σ ascertained by this fitting procedure is found to be ~20 mT throughout the inelastic tunneling regime. In comparison, σ ~5 mT is reported in the work of Koppens *et al*. for a DQD fabricated from a GaAs/AlGaAs hetero-structure grown on a standard (001) substrate [36]. Assuming the QDs in their work and ours are comparable in size (N~$10^6$), this would suggest the electronic g-factor is lower for our DQD fabricated from a GaAs/AlGaAs QW grown on a (110) substrate by about a factor of four, i.e., $g^*$~0.1 [taking the magnitude of the g-factor relevant for the (001) substrate case to be ~0.4]. The reduced electron g-factor in our case is reasonable and consistent with the value determined by a photoluminescence technique reported by Hübner *et al*. of 0.16 in magnitude for 12 nm wide (110)-oriented, GaAs/Al$_{0.32}$Ga$_{0.68}$As single QW with B∥$[\bar{1}10]$ [21]. Although the g-factor for a QD is expected to be predominately determined by the g-factor for the QW through selection of the well width at growth, aside from anisotropy with respect to applied B-field direction, there can be some modification due to the imposed lateral confining potential and charge disorder in the hetero-structure [47, 48].

In summary, we have estimated the electron g-factor for a QD circuit fabricated from a GaAs/AlGaAs QW structures grown on a (110) GaAs substrate. The in-plane g-factor of absolute value ~0.1, for B parallel to the direction [$\bar{1}$10], determined from the profile of the leakage current near zero field due to blockaded electron spin mixing with the nuclear spin of the host material via the hyperfine interaction is consistent with that reported in a work whereby the g-factor is determined by optical means for a similar QW structure [21]. We believe the report here describes the first demonstration of a gate defined QD circuit, also with charge detection capability, on a (110) GaAs substrate. Our work gives motivation for future improvements in the quality of GaAs-based materials grown on (110) substrates, namely to raise the mobility and ameliorate the current tendency for potential fluctuations leading to the formation of incidental QDs, as well as charge instabilities such as telegraph noise or drift in the charge stability diagram due to random charge trapping events. Our demonstration is a necessary step towards employing QWs grown in low symmetry directions such as [110] that offer desirable g-factor properties for electrons, and potentially holes, for applications such as coherent photon-to-spin conversion and photo-spin manipulation [49, 50]. The spin-orbit interaction anisotropy offered by (110)-oriented GaAs quantum well for quantum dot circuits is a further avenue of exploration.

We thank Sergei Studenikin for valuable contributions to this work. This work was supported by JSPS KAKENHI Grant Number JP17H06120, International Joint Research Promotion Program Osaka University, CREST JST (JPMJCR15N2); by JST [Moonshot R&D] [Grant Number JPMJMS2066]; by Asahi Glass Foundation; and by the Dynamic Alliance for Open Innovation Bridging Human, Environment and Materials. We acknowledge financial support from the grants DFH/UFA CDFA05-06, DFG TRR160, DFG project 383065199, and BMBF QR.X KIS6QK4001. This was work was also supported by the Quantum Sensors Challenge Program at the National Research Council of Canada.

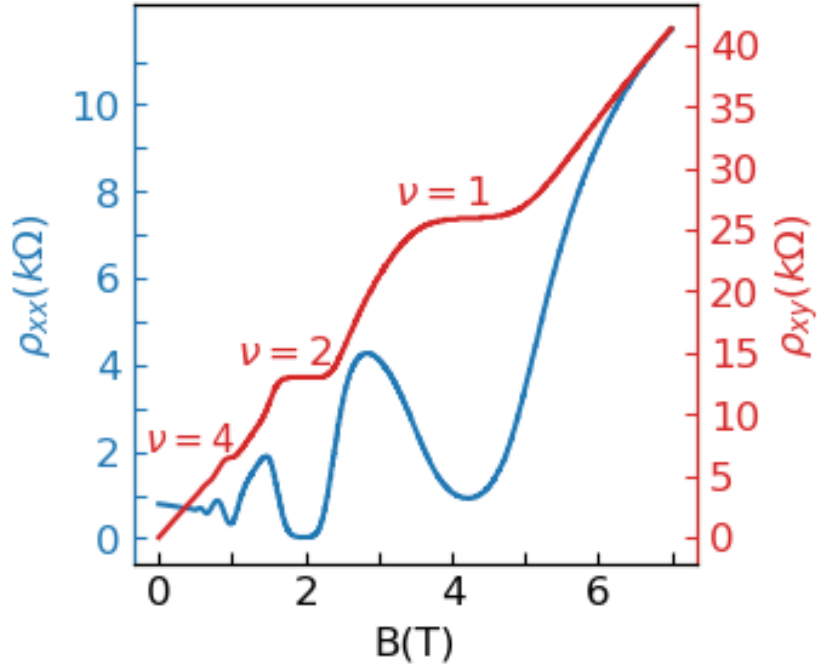

FIG. 1. Quantum Hall effect characteristics of the (110)-oriented GaAs QW. The longitudinal resistivity ($\rho_{xx}$) and the transverse resistivity ($\rho_{xy}$) are measured in the dark at 1.5 K. $\rho_{xx}$ at 0 T is 790 Ω. The carrier density and mobility respectively are $9.3 \times 10^{10}$ cm$^{-2}$ and $8.5 \times 10^{4}$ cm$^{2}$/Vs. The Ohmic contacts here are annealed at 450°C for 2 minutes. Plateaus in $\rho_{xy}$ at filling factors ν=1, 2, 3, and 4 are indicated.

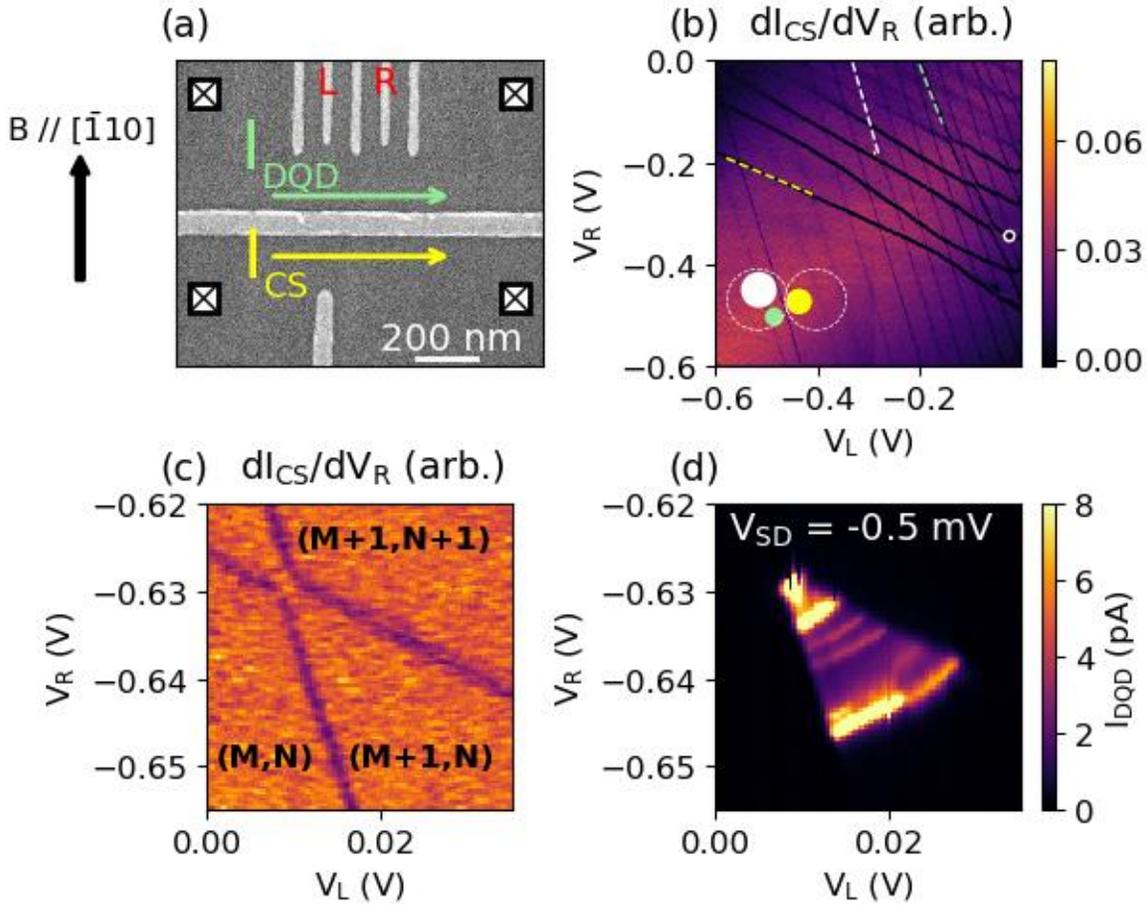

FIG. 2. (a) Scanning electron micrograph of DQD device showing active gates on the surface of the (110)-oriented GaAs/AlGaAs QW structure. A global top gate on an insulating material is also present (not shown). Voltages $V_L$ and $V_R$ respectively on the left (L) and right (R) plunger gates control the number of electrons M and N on the left and right QDs. The crossed boxes represent Ohmic contacts used to measure the current, $I_{DQD}$, through the DQD, and the current, $I_{CS}$, through the QPC charge sensor. An in-plane B-field is applied parallel to the $[\bar{1}10]$ direction. Scale bar: 200 nm. (b) Stability diagram measured with the QPC charge sensor. Here $V_{SD}$=0 mV. Cartoon inset depicts left QD region split into a larger QD (white circle) and a smaller QD (green circle) and a single QD (yellow circle) in the right QD region. The colored dashed lines with different slopes identify charge transitions in each of the three QDs. The region of the stability diagram that we subsequently focus on is indicated by the open white circle on the right. Note that after capturing the global stability diagram, and before starting the detailed examination of the target region, there was some drifting in gate voltage of the charge transition lines, and also some minor adjustment of gate voltage was necessary to measure the current. (c) Charge detection signal, $dI_{CS}/dV_R$, in arbitrary units, and (d) transport signal, $I_{DQD}$, as a function of $V_L$ and $V_R$ in the vicinity of the triangles of interest at zero B-field. Here $V_{SD}$=-0.5 mV (electrons flow from the left side to right side of DQD). The number of electrons on the left and right QDs in the various regions around the triangles are indicated in parentheses.

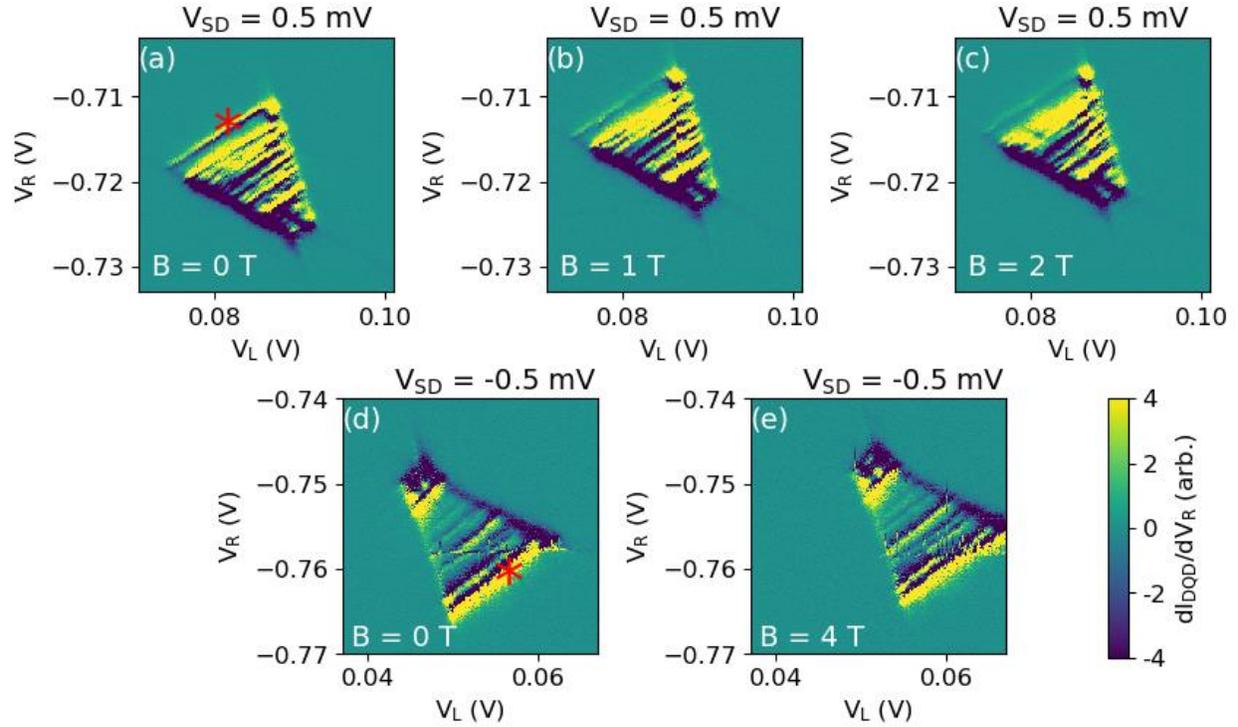

FIG. 3. (a)-(c) Transport triangles plotted as $dI_{DQD}/dV_R$, in arbitrary units, in forward bias ($V_{SD}$=+0.5 mV) showing a clear current feature along the base (indicated by asterisk) at 0 T (spin blockade is lifted due to mixing with nuclear spin), and a suppressed current feature due to spin blockade along the base at 1 T and 2 T. (d)-(e) Corresponding transport triangles in reverse bias ($V_{SD}$=-0.5 mV) showing clear current feature along base at 0 T and 4 T (spin blockade is absent). The transport triangles in some of the panels are impacted by telegraph noise and jumps in the charge stability diagram due to random charge trapping events in the material.

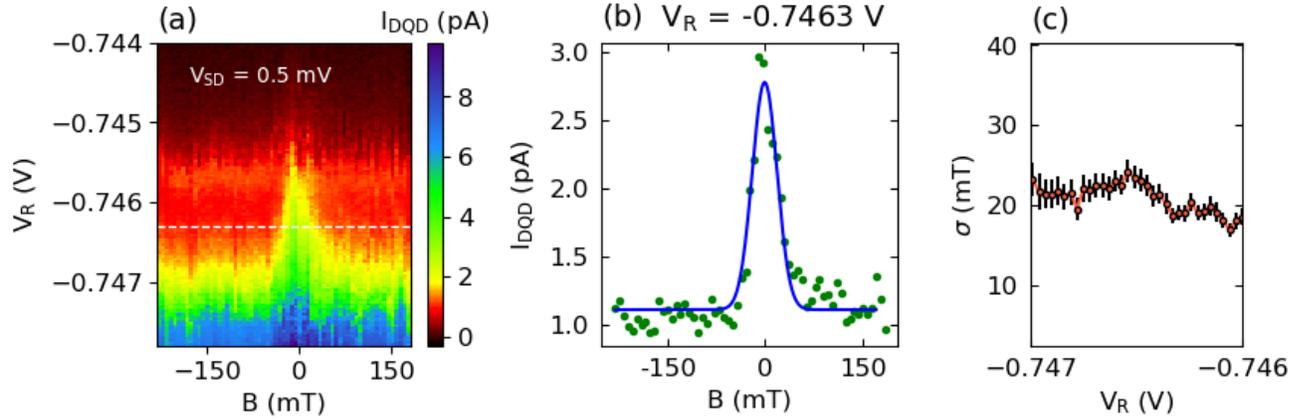

FIG. 4. (a) Detuning dependence of $I_{DQD}$ along a section through the transport triangles (at the center of, and perpendicular to, the base of the triangles) in forward bias ($V_{SD}$=+0.5 mV) for a B-field stepped in small increments through zero field. The elevated signal (leakage current) near 0 T reflects the lifting of spin blockade due to mixing with nuclear spin. The raw data has been shifted by a small amount along the B-field axis to account for a small magnetic field offset. (b) Typical $I_{DQD}$ versus B cross-section, indicated by dashed line in (a), in the inelastic tunneling regime at detuning corresponding to $V_R$=-0.7463 V and Gaussian fit giving a standard deviation σ=19±1 mT. (c) σ versus detuning throughout the inelastic tunneling regime: σ is ~20 mT over the gate voltage range shown.